\documentclass[review]{elsarticle}
\usepackage[utf8]{inputenc}
\usepackage{fullpage}
\usepackage{textgreek}
\usepackage{lineno,hyperref}
\usepackage{graphicx, float}
\usepackage{subcaption}
\usepackage{caption}
\usepackage{indentfirst}
\usepackage{siunitx}
\usepackage{amsmath}
\usepackage{gensymb}
\usepackage{booktabs}
\usepackage{multirow}
\usepackage{commath}
\usepackage{enumerate}
\usepackage{enumitem}
\usepackage{cleveref}
\usepackage{nomencl}
\usepackage{amssymb}
\usepackage{etoolbox}
\makenomenclature

\usepackage{xcolor}
\usepackage{multirow}
\usepackage[flushleft]{threeparttable}
\usepackage{textcomp}
\usepackage{gensymb}
\usepackage[export]{adjustbox}
\usepackage{tikz}
\usepackage{tablefootnote}
\usepackage{amsmath}
\usepackage[acronym,style=long3col]{glossaries}
\usepackage{chemfig}

\biboptions{sort&compress}
\begin{document}

\begin{frontmatter}

\title{Combining Transport of Pendular Water with Wind-Assisted Interfacial Evaporation for Dewatering of Concentrated Slurry Waste}

\author[mainaddress]{Tanay Kumar}

\author[secondaddress]{Hongying Zhao}

\author[mainaddress]{Xuehua Zhang\corref{correspondingauthor1}}

\cortext[correspondingauthor1]{Corresponding author}
\ead{xuehua.zhang@ualberta.ca}

\address[mainaddress]{Department of Chemical and Materials Engineering, University of Alberta, Edmonton, Alberta, Canada}
\address[secondaddress]{BC Research Inc., Richmond, BC V6V 1M8, Canada}

\begin{abstract}

Drying concentrated slurry waste is slow, particularly due to the entrapment and limited accessibility of water entrained between the particles in the slurry. A sailboat evaporator with a root-like structure is a new system that enables wind-assisted interfacial evaporation of concentrated particle slurries. In this work, we create access to the disconnected water pockets in concentrated slurry waste, facilitating faster water conduction and efficient evaporation at extremely high solid concentration. The evaporator's long roots effectively extracted water beneath 150-cm deep supernatant water layer. Through replantation of the evaporator to a separate location, an impressive evaporation rate (ER) of 4~kg/(m\(^{2}\)h) close to 80 wt\% solid concentration, a 25\% increase to a non-replanted sample. Furthermore, long periods of efficient of evaporation was achieved even at high solid concentration through hydrodynamic flushing of roots. Outdoor experiments achieved substantial volumetric reduction, yielding dried residues with over 75~wt \% solid concentration. These results underscore the system's reliable performance against highly concentrated slurries, yet to be by conventional industrial methods, including flocculation and tail-lift drying. The integration of renewable energy coupled with efficient enhancement techniques makes the sailboat evaporator a scalable and sustainable pathway for industrial wastewater dewatering.

\end{abstract}

\end{frontmatter} 

\section{Introduction}

With increasing wastewater generation and higher energy demands, interfacial evaporation has garnered research interest owing to its high energy efficiency, complete reliance on renewable energy and requirement of minimal infrastructure \cite{dao2020recent, tao2018solar, chen2025preparation, zhao2020materials, meng2021nano, zeng2023evolution,ndagijimana2024advances}. This technology has displayed efficient desalination across all salt concentrations, from low concentration seawater to high concentration brines, presenting a cost effective, sustainable way of producing high-quality freshwater. \cite{xiong2025unlocking, zhang2025efficient, wen2025nanofibrous, irshad2023highly, arshad2023exploring, asghar2025functionalized, dixit2022application, lv2022novel, wang2021integrated, sathyamurthy2023influence}. Until recently, the interfacial evaporation performance assessment has been challenging against particle-laden slurry waste, produced by various industries.

In particular, the volumetric reduction of wastewaters generated from mining industries are of great significance. It is necessary to not only expedite land reclamation timelines but also effectively store the fresh wastewaters within a limited volume \cite{wang2025influence, paulsen2025geochemical}. Current dewatering techniques include chemical flocculation, mechanical dewatering, centrifugation, and tail-lift drying \cite{feng2025high, wang2025influence, wang2023review, gumfekar2018novel, vedoy2015water, ahmed2025assessing, small2012adsorption, proskin2010freeze, kabwe2025measurement, masliyah2011handbook, li2022impact, fitton2013filtered}. They, however, struggle with these fine-particle slurries due to strong cohesion within its composition, trapping the water molecules \cite{fair2012collaboration, hyndman2010oil}, especially at high solid concentrations. Furthermore, these techniques are often hindered by high energy demands, significant operational costs, fouling issues, and restricted throughput, ultimately falling short of reaching the 75 wt\% solids concentration required for efficient dewatering and simplified disposal. \cite{zhu2017dual, revington2018process}. This highlights the pressing need for sustainable and efficient solutions to minimize the volume of these accumulating wastewaters. \cite{tusupbaev2020stimulation, morrison2022tailings}.

Interfacial evaporation enhancement for brine solutions has been extensively researched, with 3D configurations, hydrogel structure incorporation and water filtration systems \cite{irshad2023advances, he2025novel, zhang2021designing, gao2021hollow, hou20233d}. The increased efficiency in display usually combines both solar irradiation and convective flows, with the wind reducing localized vapor pressure and humidity \cite{choi2025simultaneous, li2025wind, li2021solar, zhang2023distinct, liu2022evaporation}. For instance, Chen et al. reported a water evaporation rate of up to 11.9 kg/(m\(^{2}\)h) for a hollow, open photothermal evaporator under 1.0 sun irradiation with a 4.0 m/s wind  \cite{chen20223d}. Wind-assisted interfacial evaporation is especially relevant to high-latitude regions, where reduced solar irradiation and lower ambient temperatures demand the utilization of renewable wind energy to enhance convective vapor transport. \cite{choi2025simultaneous, li2025wind, irshad2025controlled, philip2013performance, li2022magnetic, wu2023boosting,palimi2025enhanced}. 

The interfacial evaporation of particle-laden slurries, however, present fundamentally distinct challenges from that of saline water. Unlike brine solutions, where water conduction remains relatively consistent, slurry systems exhibit highly variable water transport behavior, dictated by the composition, viscosity and solid concentrations of the suspension. At very low concentration, water exists as a continuous phase between the particles. However, at this pendular regime, the water is in strong contact with the particles, with air as the continuous phase \cite{koos2011capillary, gögelein2010controlling}. In our previous study, we investigated a root-inspired evaporator system for the solar-driven volumetric reduction of fine-particle suspensions \cite{kumar2024root}. Building upon this, we subsequently developed a wind-optimized evaporator design, which demonstrated an evaporation rate exceeding 10 kg/(m\(^{2}\)h) for concentrated silica slurry \cite{kumar2025wind}. We also investigated the influence of different root depths, structures and slurry compositions on the wind-assisted evaporation behavior, especially at the high solid concentration regime. 

The evaporation enhancement of the slurry in the pendular regime remains unexplored. In this study, we address these gaps by implementing multiple strategies aimed at accelerating the evaporation of high solid concentration wastewaters. Specifically, we explore the influence of suspension viscosity on evaporation performance, assess how replantation of the evaporator affects spatial moisture distribution, and evaluate performance enhancement through suspension flushing. Additionally, we investigate the role of mechanical vibration on the efficiency of our evaporator. We also provide a theoretical analysis of the effect of the enhancement techniques at this high solid concentration regime, complemented by comprehensive experimental characterization. Finally, we demonstrate the evaporator's versatility at multiple scales, covering diverse tailing pond conditions with outdoor evaporation experiments of over 20 L of wastewater suspension, unequivocally showcasing the scalability of our evaporator setup. These findings demonstrate the effectiveness of the proposed strategies in accelerating slurry evaporation within the pendular regime using interfacial evaporation systems, an outcome unattainable through conventional methods.

\section{Material and methods}

\subsection{Chemicals and materials}

Fine silica suspensions were prepared by mixing varying proportions of fine particles (IMCD IMSIL A-10, median diameter: 2 μm) in tap water to achieve the desired solid concentrations for each experiment. Bitumen samples used in the study were sourced from an industrial supplier. Additionally, real fluid fine tailings were procured from an industrial site. These tailings typically consisted of 65–70 wt\% water and 25–30 wt\% solids, which included 20–25 wt\% clay, 3–5 wt\% sand, and 2–5 wt\% silt. The mixture also contained 1–3 wt\% bitumen, less than 1 wt\% salts, and trace levels (\textless0.01 wt\%) of naphathenic acids, solvents, and other industrial chemicals.

\subsection{Experimental setup}

\begin{figure}[H]
  \centering
  \includegraphics [width = 16.5 cm]{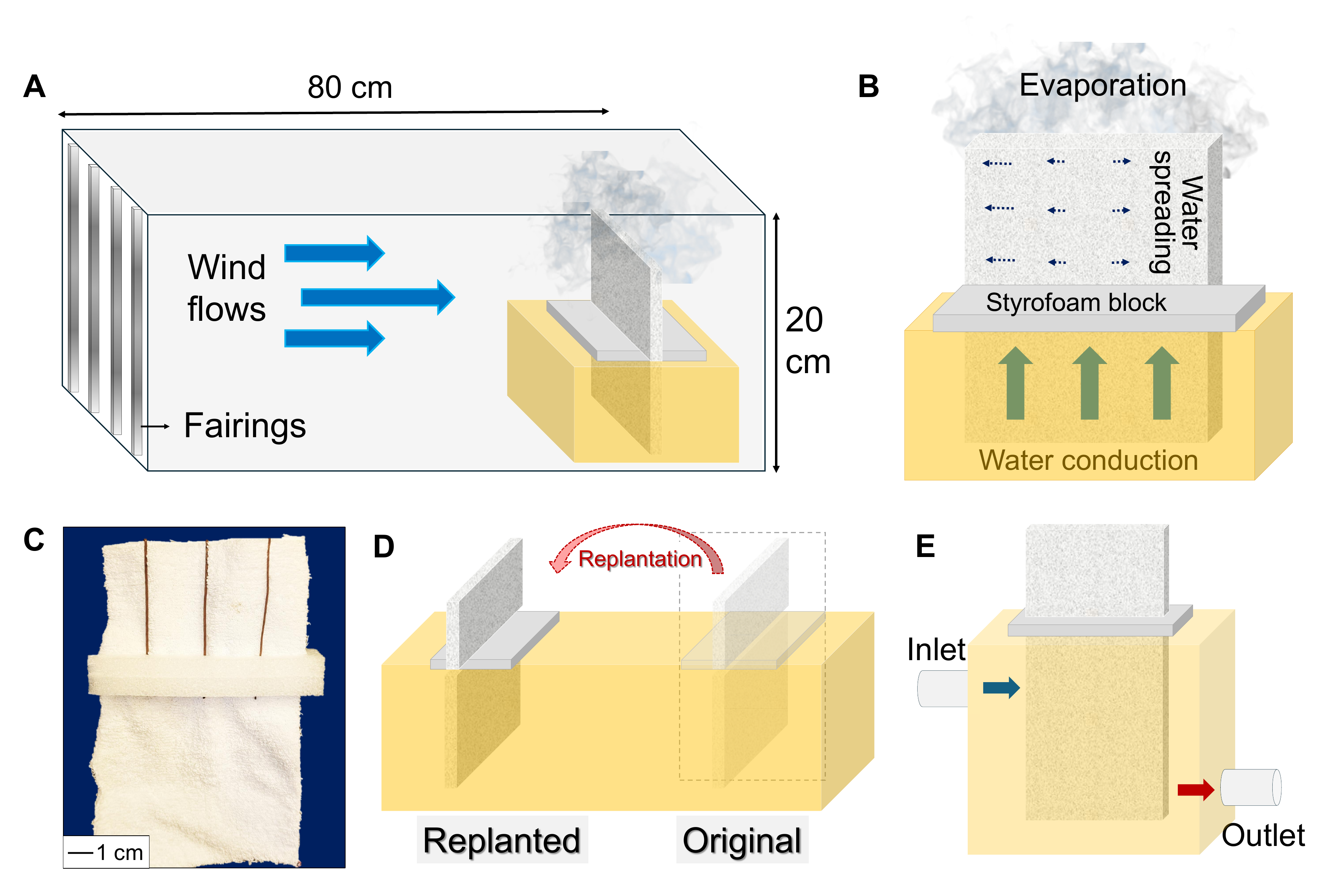}
  \caption{A) Sketch of the wind tunnel for the wind evaporation experiments; B) Schematic of the sailboat evaporator (SE) setup, with illustration of the various physical processes; C) Image of the SE setup; Schematic diagram of evaporation enhancement through D) sailboat replantation to a new location and E) hydrodynamic flushing. }
  \label{fig1}
\end{figure}

For the evaporation experiments, a custom wind tunnel, using rigid cardboard and transparent acrylic sheets, was constructed to provide a stable, laminar airflow, as illustrated schematically in Fig. \ref{fig1}(A). The sail of the evaporator was placed perpendicular to the direction of wind, originating from a fan placed at the opposite end of the tunnel. The fan delivers wind velocities in the range of 0-4 m/s, corresponding to calm to gentle breeze conditions (Beaufort Scale \textless 3). To ensure laminar flow conditions, a fairing was installed adjacent to the fan inlet. Laminar flow was validated by confirming a uniform wind velocity distribution across the tunnel’s outlet, within a permissible variation of ±10 \%.

The sailboat evaporator comprises of two main components: (1) a mainsail structure positioned at the top to capture wind flow, going all the way down inside the suspension (roots) to draw water and (2) an insulating foam base that both supports the evaporator and maintained buoyancy, as shown in Fig. \ref{fig1}(B). A photographic image of the sailboat evaporator is presented in Fig. \ref{fig1}(C). For the sailboat setup, the dimension of the mainsail was fixed at 7 cm in height and 11 cm in width. For the dendritic roots structure, additional fibers are stitched together and distributed throughout the slurry matrix. The dendrites root structure feature the same total contacting area but wider spreading in the slurry volume.

Different setups for evaporation enhancement at high slurry concentration are displayed in Fig. \ref{fig1}. In the replantation technique, the evaporator setup is placed from one side of the container and replaced to the other side (Fig. \ref{fig1}(D)). For the suspension flushing technique, a container with an inlet and an outlet is utilized to allow the introduction of fresh and the removal of old suspension ((Fig. \ref{fig1}(E)). While the SE setup can also be utilized for the collection of the evaporated water, our work focuses on the volumetric reduction of these suspensions.

\subsection{Characterization of evaporator setup}

The volumetric reduction of the particle-laden suspension was determined by water loss from evaporation, leading to increasing solid concentration. The weight loss within the suspension is measured over time using an electronic scale, measuring upto 50 kg with an accuracy of 0.1g. The evaporation rate (ER) was calculated by calculating the mass loss rate, followed by normalizing with the evaporative surface area. The experiments were performed in a laboratory environment, with the relative humidity of 30 \% and an ambient temperature of 20 $^{\circ}$C. The moisture content distribution throughout the suspension was measured using a 2-pin hygrometer, with the pin completely immersed inside the slurry. The wind speed for the experiments indoors and outdoors were recorded using a handheld anemometer (WUYUZI Anemograph), with a wind velocity accuracy of ±3 \%. Temperature measurement was performed by placing a handheld thermography camera (Hikmicro B series) 15 cm away from the setup, in accordance with the setting instructions of the device for accurate measurement.

To visualize the drying process, a Hele-Shaw Cell was constructed with two thin glass slides (10 cm x 10 cm) separated by a 1 cm thick spacer. The image of the setup is present in the Supporting Information. The cell was kept intact using clips, ensuring there is no space in between the glass slides for leakage. The cell was kept upright, filled with 50 wt \% silica slurry and the root was dipped inside the slurry. The 3-D image of the silica slurry was obtained through Micro-CT X-Ray imaging (ZEISS Xradia Versa 620 X-ray Microscope, Zeiss, Germany). The suspension is placed on top of a sample holder of 2 cm diameter and is placed upright inside the X-Ray chamber. The different phases are identified by the gradient on the gray scale owing to the difference in the densities of the constituents in the composition. The viscosity of the suspensions was calculated using a rheometer (Netzsch). 

The water getting transported to the sail surface is measured using FTIR spectroscopy. A piece of the cotton sail is cut from different parts of the sail and placed under the probe for measurement. The particle fouling on the sail and the root of the setup is measured using a handheld Raman spectrometer. The influence of silica flushing on a fouled root is measured using a flushing loop setup with 10 wt\% silica used as the flushing liquid. For the vibrational experiment, the angle of meniscus within the capillary tube is measured using a contact angle meter. The silica particle coverage on the root surface is captured using a digital camera and analyzed using ImageJ software.

\section{Results and Discussions}

\subsection{Influence of slurry viscosity on evaporation rates}

\begin{figure}[H]
  \centering
  \includegraphics [width = 15.5 cm]{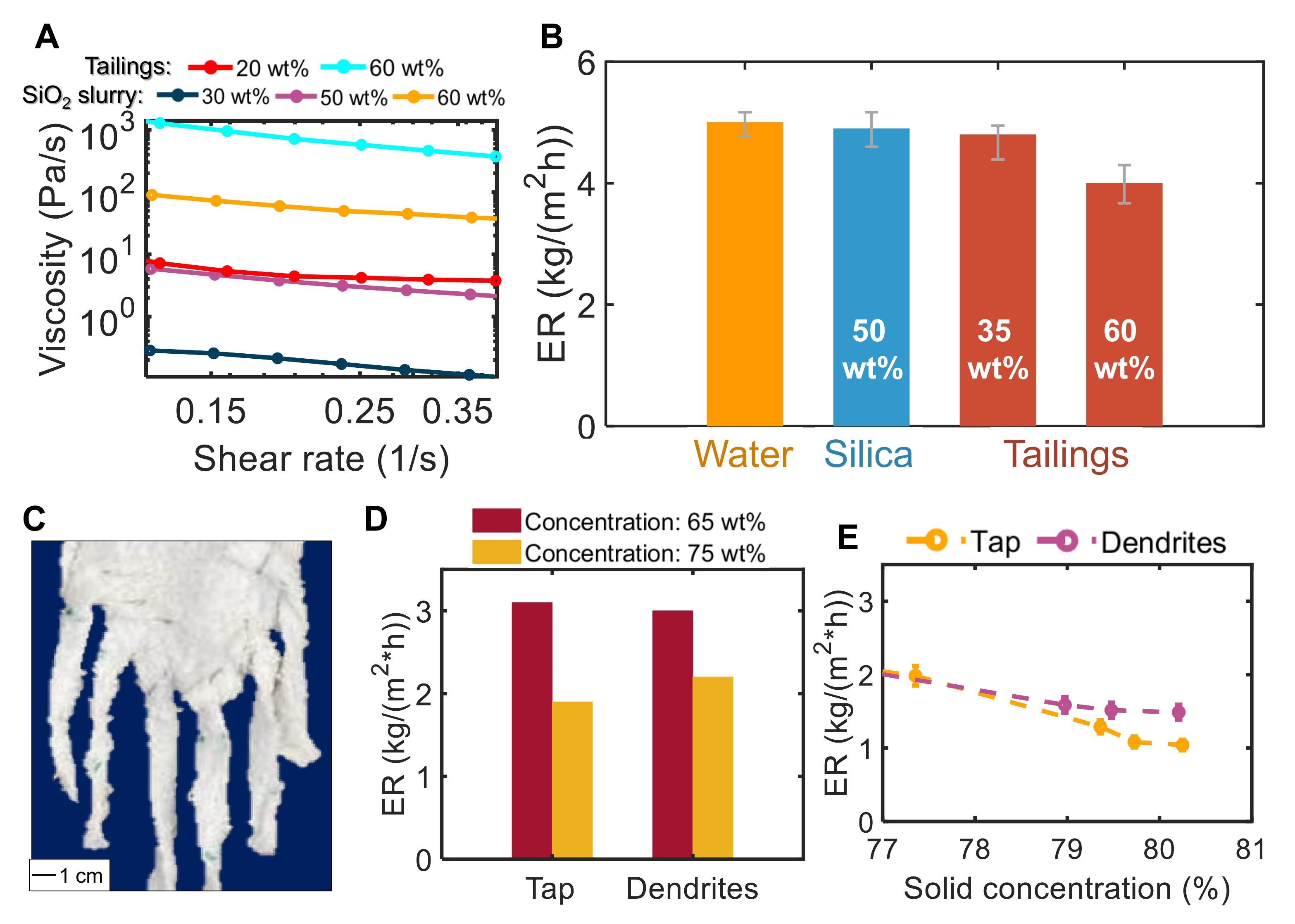}
  \caption{A) Viscosity vs shear rate of different particle-laden suspensions; B) ER comparison between water, silica and tailing suspensions; C) Image of the dendritic root structure; D) ER vs the root structure beyond 65 wt\% and 75 wt\% solid concentration; E) ER vs solid concentration of tap and dendritic roots setup beyond 77 wt\%. }
  \label{fig2}
\end{figure}

The influence of the solid concentration and composition on the viscosity of the particle slurry is evaluated. For silica slurry, an increase in the solid concentration led to an increase in the viscosity, rising from 0.3 Pa/s for 30 wt\% to 90 Pa/s for a 60 wt\% slurry (Fig. \ref{fig2}(A)). A similar trend is observed for tailing wastewaters, with the viscosity rising from 8 Pa/s to 900 Pa/s as the solid concentration increased from 20 wt\% to 60 wt\%. Additionally, the tailing composition displays nearly 10 times higher viscosity than the silica fines suspension of same solid concentration. This is attributed to the presence of clay particles and hydrophobic bitumen content within the tailing composition, enhancing particles cohesion and suspension viscosity.

The ER of water is evaluated against different particle suspensions using the sailboat evaporator setup at a wind velocity of 4 m/s. As displayed in Fig. \ref{fig2}(B), the highest ER was displayed by water (5.1 kg/(m\(^{2}\)h)), slightly higher than 50 wt\% silica slurry (4.9 kg/(m\(^{2}\)h)). On the contrary, the ER was lower for tailing suspensions, reducing from 4.6 kg/(m\(^{2}\)h) for 35 wt\% to 4 kg/(m\(^{2}\)h) for 60 wt\% solid concentration. A little reduction in ER is observed as the solid concentration of silica fines goes from 0 wt\% to 50 wt\%, while a more significant change is observed for the tailings suspension going from 35 wt\% to 60 wt\%. 

\subsection{Evaporation enhancement at high solid concentration}

\subsubsection{Influence of root structure on evaporation rates}

The influence of addition of dendritic roots on the ER of the setup is evaluated and compared against a tap root setup, as displayed in Fig. \ref{fig2}(C). A silica slurry of 65 wt\% initial solid concentration and a wind flow of 2.8 m/s is utilized. At a solid concentration of 65 wt\%, the efficiency of both setups is similar, displaying an ER of around 3 kg/(m\(^{2}\)h) (Fig. \ref{fig2}(D)). The results suggest little influence of the dendrites addition at that solid concentration regime. On continued evaporation, the ER steadily decreases as the solid concentration goes beyond 75 wt\%. However, the dendrite setup displays slightly higher ER (2.2 kg/(m\(^{2}\)h)), \textgreater20\% larger than the tap root setup (1.9 kg/(m\(^{2}\)h)). As depicted in Fig. \ref{fig2}(E), a magnified view of the ER reveals the dendritic root setup sustaining the evaporation to a higher solid concentration (1.5 kg/(m\(^{2}\)h)), exhibiting a less steeper slope than its tap root counterpart (1 kg/(m\(^{2}\)h)). At very high slurry concentration, the incorporation of dendrites in the root structure within the evaporator sufficiently enhanced the evaporation performance.

\subsubsection{Replantation of the evaporator setup}

\begin{figure}[H]
  \centering
  \includegraphics[width=16.8 cm]{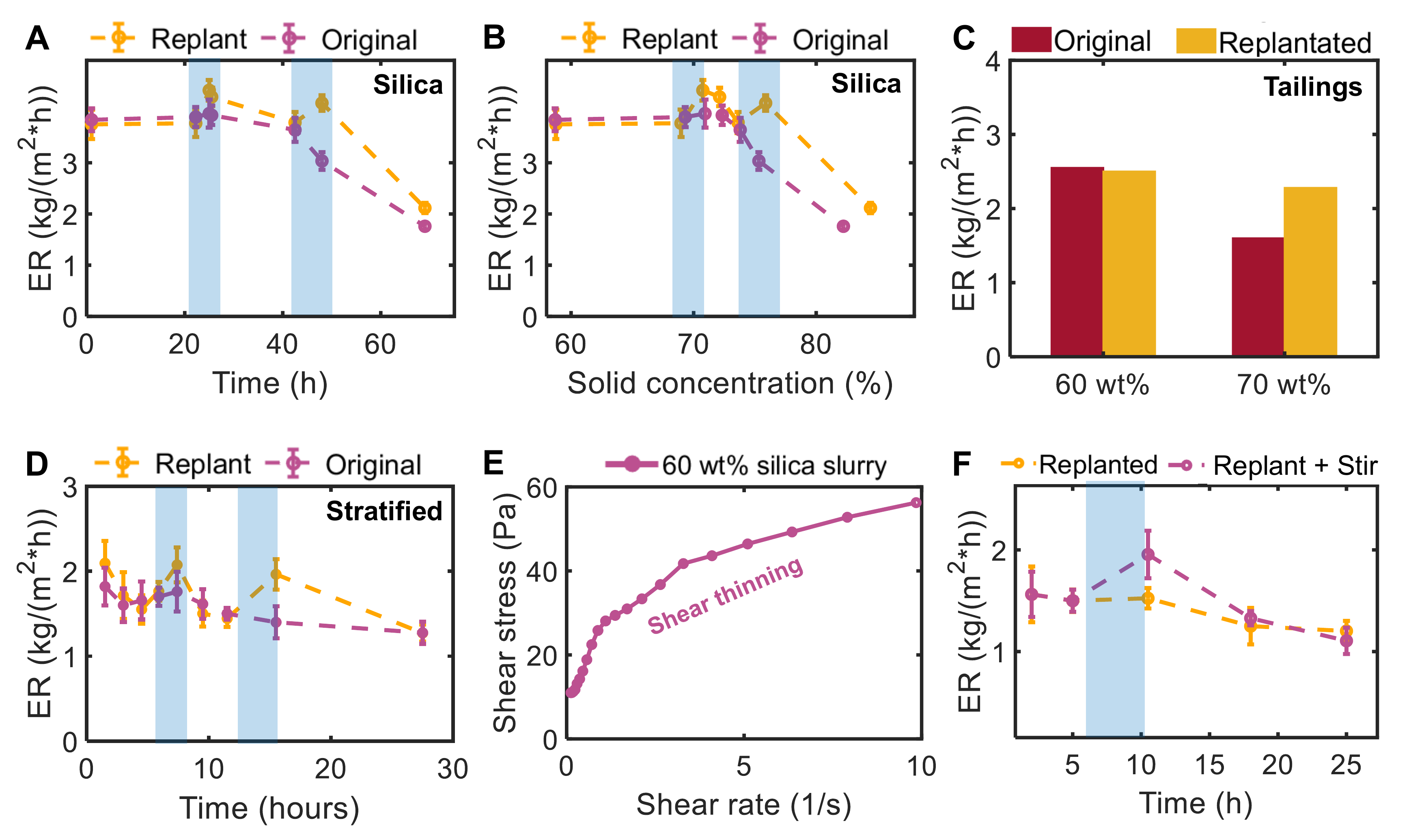}
  \caption{
 a) ER vs A) time and B) solid concentration exhibiting the effect of replanting the SE setup into a new location against 60 wt\% silica slurry (blue zone: replantation induced); C) ER vs solid concentration of tailing suspension exhibiting the effect of replanting the SE setup; D) ER vs time displaying the effect of replanting for a stratified 60 wt\% tailing setup; E) Shear stress vs shear rate for a 50 wt\% silica slurry; F) ER vs time exhibiting the influence of stirring on the replantation process (blue zone: stirring induced).
  }
  \label{fig3}
\end{figure}

To enhance the ER at high solid concentration, the SE setup is physically replanted to a different location within the particle suspension. The evaporator is tested against a silica suspension of 65 wt\% solid concentration under a wind speed of 3.5 m/s. As exhibited in Fig. \ref{fig3}(A), initially, the ER of both the setups is similar. When the evaporator was replanted to a different location after 20 hours, a surge in the ER is observed, leading to a 10\% increase in ER. On continued evaporation, there is a reduction in ER for both the setups, with a steeper decrease in the replanted setup. When the evaporator was replanted again, a bigger surge in ER is observed (4.2 kg/(m\(^{2}\)h)), with a near 40\% jump as compared to the the non-replanted setup (3 kg/(m\(^{2}\)h)).

The ER of the both the setups is further evaluated against slurry concentration (Fig. \ref{fig3}(B)). The first replantation is observed at around 70 wt\% solid concentration, while the second replantation occurs beyond 75 wt\%. The increase in ER during the first replantation is attributed to access to additional slurry-bound water in the new location. After the second replantation, the increase in ER is much larger than the first. This is credited to a shift in slurry behavior beyond 75 wt\% from thin, continuous water channels to discontinuous water clusters, which are inaccessible by the evaporator setup \cite{kumar2024root}. During removal of the roots, high density of silica particles are present on the root surface, with little shearing of the silica particles during the replantation procedure. For tailings, similar ER is observed at 60 wt\% (Fig. \ref{fig3}(C)). However, at 70 wt\%, a 30\% increase in the ER is observed upon replantation, rising from 1.6 kg/(m\(^{2}\)h) to 2.28 kg/(m\(^{2}\)h).

We next simulate a two-layer stratified system consisting of a 50 cm supernatant water layer on top and the concentrated slurry (\textgreater65 wt\%) in the bottom, with and without SE setup replantation. The root structure is immersed exclusively in the concentrated slurry, while the surrounding supernatant layer is isolated using a plastic covering, preventing any direct contact. The setup is pre-wetted and the ER is evaluated at a wind velocity of 2.8 m/s. As observed in Fig. \ref{fig3}(D), the ER of both setups remains similar initially, around 1.8 kg/(m\(^{2}\)h). Upon replantation, a surge in ER is observed, increasing the evaporation efficiency by 10\%. A second replantation at a higher solid concentration, results in a bigger surge in ER, increasing by nearly 33\% from 1.5 kg/(m\(^{2}\)h) to 2 kg/(m\(^{2}\)h). This transitory increase in ER value aligns well with the previous replantation results.

The non-uniform moisture distribution along with the disconnected water clusters is exploited through the replantation process. The rise in ER, however, is brief. Post replantation, the proximal water pockets are immediately conducted by the evaporator, beyond which a sudden reduction in ER is observed. Stirring the suspension at high solid concentration may allow homogenization of the particle suspension. On shearing a 50 wt\% silica suspension, a non-linear increase in shear stress is observed with increasing shear (Fig. \ref{fig3}(E)). The behavior corresponds to a shear-thinning suspension, suggesting a reduction in viscosity on stirring the suspension.

The influence of replantation along with suspension stirring is evaluated at a high solid concentration (60 wt\%) under a wind velocity of 2.3 m/s. As displayed in Fig. \ref{fig3}(F), stirring the suspension improved the ER by 20\%, increasing from 1.6 kg/(m\(^{2}\)h) to 1.95 kg/(m\(^{2}\)h). This improvement in ER is attributed to homogenization of the suspension, redistributing the moisture within the matrix. However, close to 70 wt\%, the silica suspension crosses the plastic limit, transitioning from its plastic state to a semi-solid state. Stirring beyond the plastic limit is ineffective, a concentration regime where replantation is more effective.

\subsubsection{Characterization of evaporator and wastewater suspensions}

\begin{figure}[H]
  \centering
  \includegraphics [width = 10.5 cm]{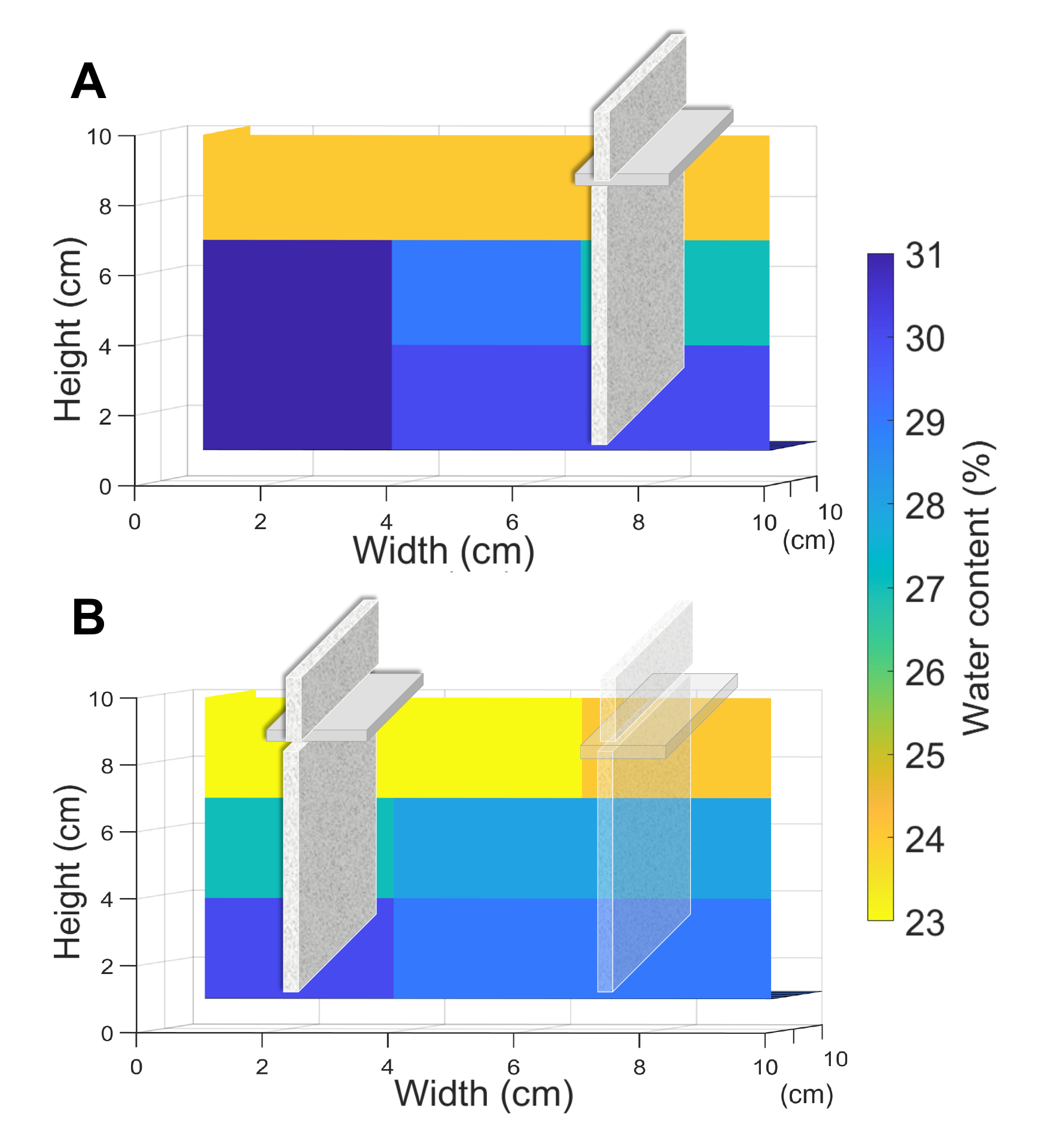}
  \caption{Spatial moisture distribution within the silica slurry after 45 hours A) without and B) with replanting the SE setup. 
}
  \label{fig4}
\end{figure}

The spatial moisture distribution within the slurry is analyzed for a replanted and non-replanted sample. The SE setup was used to evaporate 2 L of 65 wt\% silica slurry under a wind velocity of 4 m/s. The moisture distribution is measured for both the setups after 40 hours of evaporation, with replantation performed 20 hours after experiment initiation. In both the setups, moisture content increased from the root apex (low water content) to the slurry base (high water content). For radial gradient, less uniformity is observed for the non-replanted setup (variance \textgreater5\%), with a volume of slurry with higher moisture content away from the setup (Fig. \ref{fig4}(A)). Conversely, for the replanted setup, a more moisture uniform distribution is observed throughout the suspension volume, with a variance \textless3\% (Fig. \ref{fig4}(B)). 

After 40 hours of evaporation, the replanted setup achieved an average solid concentration of 76 wt\%, surpassing the non-replanted setup (72 wt\%). This disparity arises from the access of the discontinuous, isolated moisture pockets within the slurry matrix, reducing localized moisture retention. This confirms a much higher enhancement in the ER through the replantation technique, especially at a high solid concentration beyond the transition point of silica slurry.

\begin{figure}[H]
  \centering
  \includegraphics [width = 15 cm]{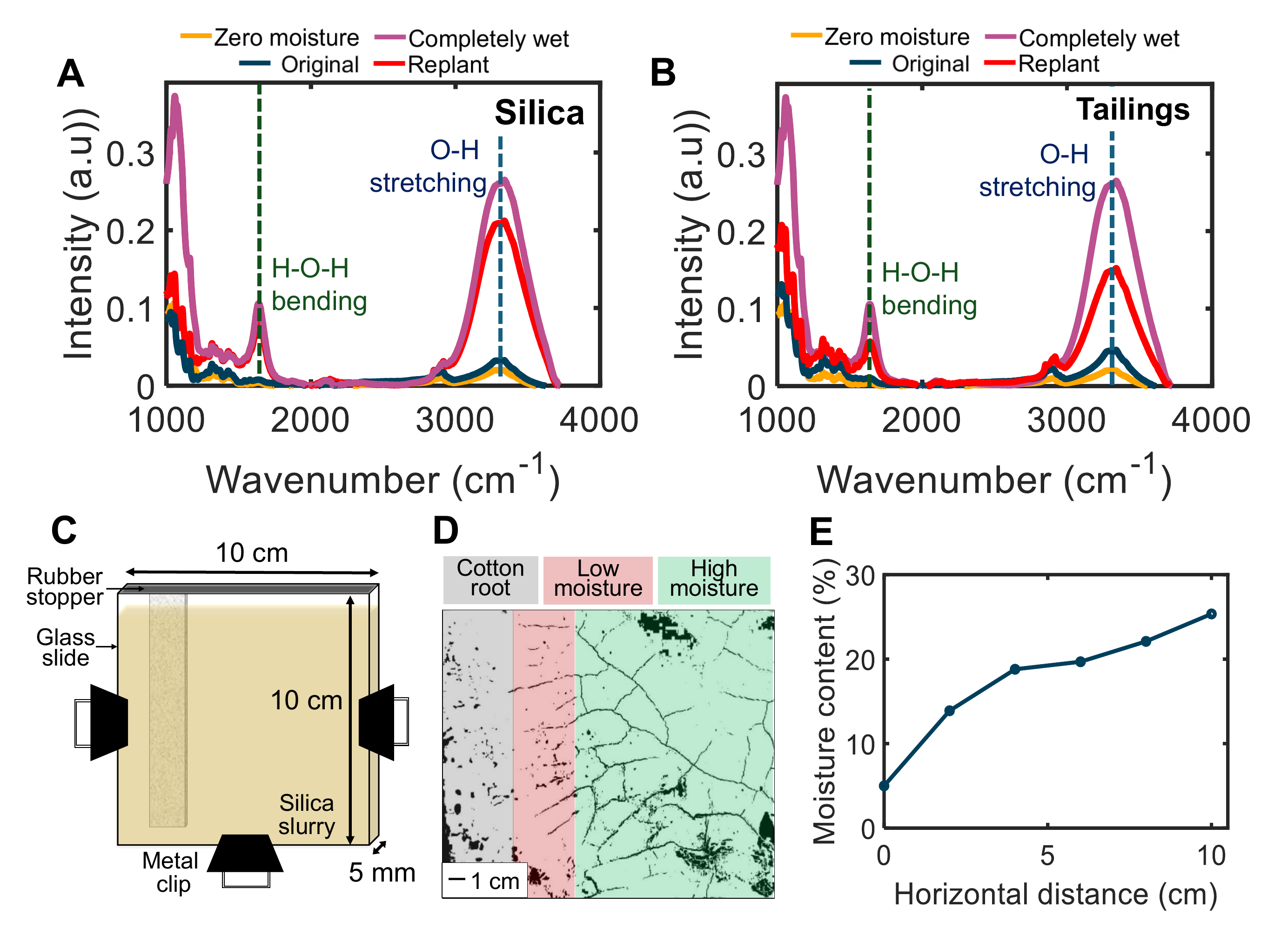}
  \caption{FTIR spectrum of water on the cotton sail surface with and without replanting for evaporating a A) silica and a B) tailing suspension; C) Image of the Hele-Shaw apparatus; D) X-Ray imaging of moisture distribution within the silica slurry though a Hele-Shaw setup; E) Moisture content \% as a function of horizontal distance away from the root in the Hele-Shaw setup. 
}
  \label{fig5}
\end{figure}

FTIR Spectroscopy is performed to detect the presence of water on the sail surface after replantation. As displayed in Fig. \ref{fig5}(A), without any replantation in silica slurries, little peaks are observed for the H-O-H bending and the O-H stretching bonds. This is attributed to little water conducted to the sail surface with continued evaporation. Conversely, after replantation, a defined, larger peak is observed, suggesting the presence of sufficient water. A similar trends is observed for tailing suspensions of 60 wt\% as well, as shown in Fig. \ref{fig5}(B), with a much larger H-O-H and O-H peak after replantation of the evaporator setup. The lower and upper-limit boundaries for the peaks are defined using dry (0\% water) and water-soaked (100\% water) cloths. This permits a rough quantification of the water present within the pores of the cotton sail. For silica slurries, 19\%`of the porous area is initially filled with water. After replantation, close to 87\% of the porosity gets filled with water. This signifies higher water conduction, which in turn leads to a higher ER. Similarly, against tailing suspensions, the amount of water in the cotton sail increases from 14\% to 62\%, thereby enhancing the ER.

The spatial water distribution within slurries after replantation during root-assisted evaporation is visualized utilizing a Hele-Shaw cell \cite{bararpour2024pattern, de2023convective} within a cabinet X-ray chamber. An evaporation experiment under 4~m/s wind velocity is conducted with a miniature evaporator setup inserted inside the Hele-Shaw cell, with the volume filled with 50~wt \% silica slurry. The sketch of a Hele-Shaw cell along with the dimensions are present in Fig.~\ref{fig5}(C). 

An X-Ray image is taken at the 80 wt \% solid concentration, as shown in Fig.~\ref{fig5}(D), with three distinct areas. On the left side of the setup is the root, from where the water absorption takes place. Well-defined cracks are observed in the suspension in the slurry region proximal to the root structure. This phenomenon is attributed to the reduced availability of free water near the root structures at high solid concentrations, driven by extreme negative matric potential disrupting porous capillary networks within the slurry \cite{luo2023effects, jabbarzadeh2024thermo}. 

Away from the root structure, fewer cracks along with isolated moisture pockets appear within the slurry. This is attributed to the lack of water conduction in areas away from the roots with the absence of the water capillaries. To quantify the inhomogeneous water distribution, the moisture content was quantified against the horizontal distance away from the roots (Fig.~\ref{fig5}(E)). For a silica slurry of average solid concentration of 80 wt\%, the moisture \% increases as we move away from the roots, ranging from 5\% just next to the roots, to 24\% at the farthest horizontal point away from the root (10 cm away). This confirms the presence of water clusters away from the evaporator setup, which are accessed by the replantation technique.

\subsubsection{Hydrodynamic flushing of the evaporator roots}

\begin{figure}[H]
  \centering
  \includegraphics [width = 14 cm]{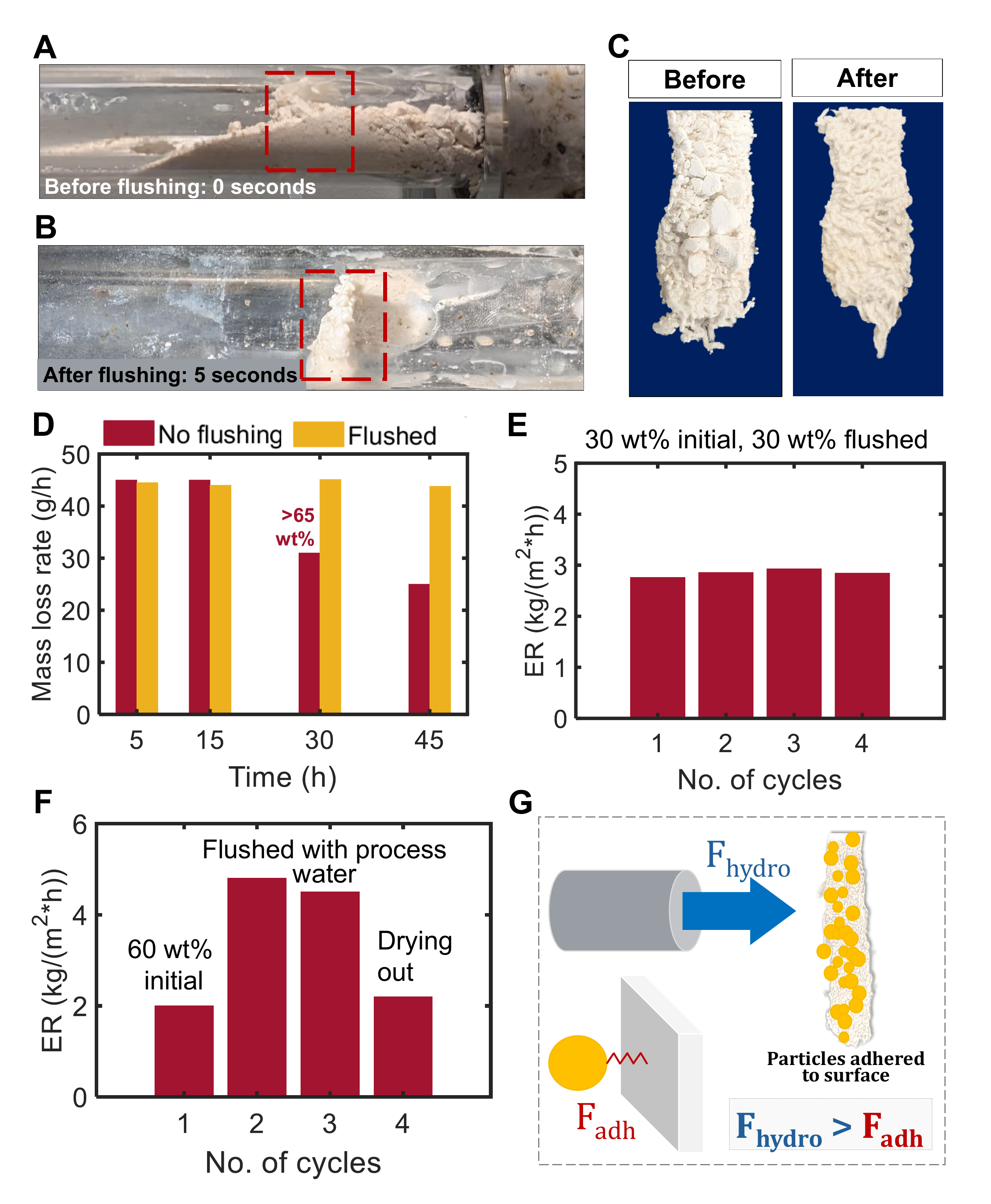}
  \caption{Image of the evaporator root in silica slurry A) before flushing and B) 5 seconds after flushing with 10 wt\% silica suspension in the loop setup (Root within the red box); C) Image of the evaporator root before and after the hydrodynamic flushing; D) Mass loss rate v/s time for the SE setup with and without silica flushing; E) Mass loss rate vs time for a 30 wt\% tailing wastewater flushed with a 30 wt\% tailings; F) Mass loss rate vs time for a 60 wt\% tailing wastewater flushed with process water; G) Schematic displaying the force balance during the flushing process.
}
  \label{fig6}
\end{figure}

Mature, concentrated tailing ponds are often refilled with freshly produced tailings transported into the ponds. This may lead to a hydrodynamic flushing process by the fresh slurry, which may enhance the evaporation efficiency, especially at high solid concentration. To evaluate, a flushing loop setup is utilized with a 10 wt\% silica suspension as the flushing liquid. We test the setup against a dried root setup completely buried in highly concentrated silica slurry, as shown in Fig.~\ref{fig6}(A). After 5 seconds of flushing under low velocity (1.6 m/s), almost all the deposition around the roots is removed (Fig.~\ref{fig6}(B)).

The effect of hydrodynamic flushing on the root fouling is optically observed. Big chunks of silica deposition is visible initially on the root surface, as displayed in Fig.~\ref{fig6}(C). Post flushing, however, the deposited silica particles are removed, with a clean surface visible. To evaluate the effect on the evaporation performance, the SE setup is tested with and without flushing against a 50 wt\% silica suspension under a wind velocity of 4 m/s. Fresh suspension of 50 wt\% are introduced periodically through the inlet with the old thicker slurry replaced out from the outlet.

The evaporator displays a high mass loss rate of 45 g/h for the first 15 hours of the experiment, irrespective of flushing (Fig.~\ref{fig6}(D)). This is because there is enough free water available to sustain high ER for both the setups. However, beyond 15 hours, a drop in the mass loss rate is observed, attributed to the slower water conduction at high solid concentration (>65 wt\%), as discussed previously. Introduction of fresh suspension, however, exhibits a high mass loss rate (44 g/h) throughout the experimental duration, resulting in a massive 75\% improvement in efficiency than its counterpart (25 g/h) at higher solid concentration. 

The solid composition of the wastewater may have an influence on the flushing dynamics. The SE setup is utilized against a 30 wt\% tailing wastewater under a wind velocity of 4 m/s, with the incoming tailings of the same solid concentration (30 wt\%). The experiment is run for nearly 40 hours through 4 cycles of 10 hours, with suspension flushing done at the end of each cycle. An average mass loss rate of 41 g/h is observed during the first cycle of evaporation, as exhibited in Fig.~\ref{fig6}(E). Post hydrodynamic flushing, the evaporator sustains the same evaporation performance throughout the next cycle. The same trend is observed over the next two cycles of evaporation, with similar mass loss rate to the first cycle. The sustenance of the mass loss rate is attributed to the removal of silica deposition on the root surface by the suspension flushing.

The solid concentration of the flushing liquid may also contribute in improving the ER of the setup. An ER is tested against tailing wastewater of 60 wt\% initial concentration under a wind velocity of 4 m/s, with hydrodynamic flushing using process water containing negligible solids (Fig.~\ref{fig6}(F)). A low average mass loss rate of 26 g/h is observed in the first cycle. This is due to the high solid content of the tailing suspension limiting the rate of water conduction. After hydrodynamic flushing after the first cycle, a massive surge in the mass loss rate is observed, increasing by 250\% to 65 g/h. This increase is attributed to the conduction of the free process water rather than the thick tailing suspension. A similar high mass loss rate is observed in the next cycle post-flushing. 

To test our hypothesis, no flushing is done before the beginning of the fourth cycle, but rather the suspension is allowed to dry out. A sharp decrease in the mass loss rate is reported, falling from 60 g/h in the third cycle to 27 g/h in the fourth. This decrease is attributed to the absence of the free process water for water conduction, but rather a complete reliance on the thick tailing suspension only. The mass loss rate in the 4th cycle is very similar to the 1st cycle, supporting our hypothesis. We conclude that the solid concentration of the flushing water also influences the evaporation behavior.

\subsubsection{Force of hydrodynamic flushing}

A particle adhered to the root surface will be removed by incoming flow if the hydrodynamic force (F\textsubscript{hydro}) exceeds the particle-root adhesive force (F\textsubscript{adh}), as shown in Fig.~\ref{fig6}(G). The relationship is defined as:

\begin{equation}
F_{hydro} \geq F_{adh}
\end{equation}

Assuming a steady-state, intermediate velocity flow (1 \textless Re \textless 800), F\textsubscript{hydro} on a singular, spherical particle by water is given by the following relation:

\begin{equation}
F_{hydro} = 0.5C_d\rho Av^2
\end{equation}

where C\textsubscript{d} is the drag coefficient, \(\rho\) is the density of the flushing liquid, A is the area of the particle and v is the velocity of the incoming liquid. As stated in the equation, the hydrodynamic flushing force is directly proportional to the density of the flushing liquid and to the square of the flushing velocity. For a group of particles, however, the particle packing may cause reduced effective fluid velocity and increased resistance, thereby influencing the hydrodynamic force generated.

\begin{equation}
F_{hydro} \propto (1- \phi)^n
\end{equation}

where \(\phi\) is the local volume fraction of particles packed on or near the fibrous root surface and n is the numerical coefficient (ranging between 2.4–3 for moderate Re). A higher solid concentration volume on the root surface would also require a higher hydrodynamic force to remove the particles off the root surface. A detailed derivation of the hydrodynamic force for a group of particles is shown in the Supporting Information.

\subsection{Effect of vibrations on the evaporation performance}

\begin{figure}[H]
  \centering
  \includegraphics [width = 16.5 cm]{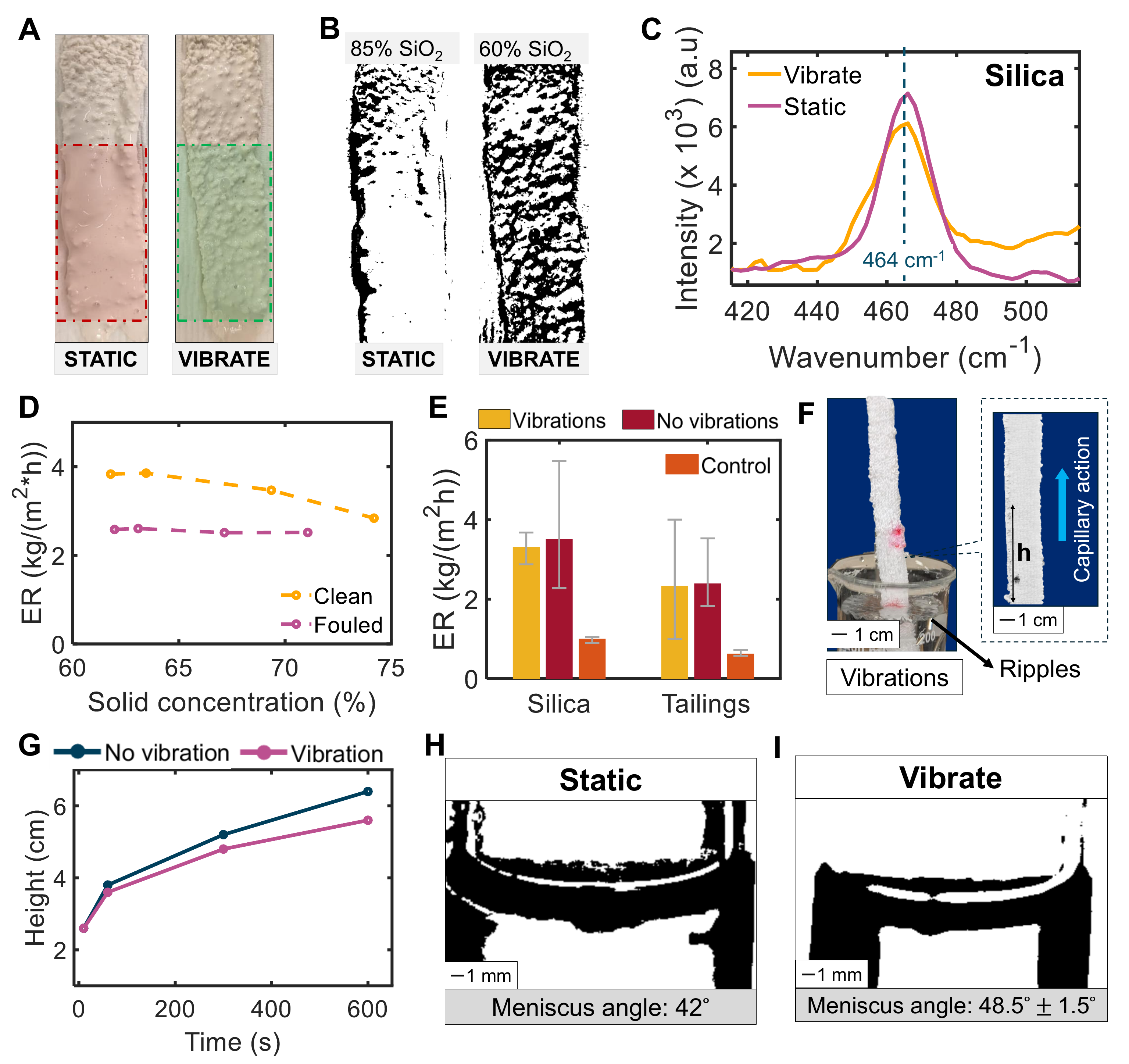}
  \caption{A) Images of the fouling on the roots with and without vibrating the evaporator root; B) Optical image processing for surface coverage of silica; C) Raman data displaying silica particles in vibrating and non-vibrating roots; D) ER vs solid concentration with a clean and a fouled root; E) ER comparison between 60 wt\% silica and tailing suspension in a vibrating and non-vibrating container; F) Schematic of the capillary wicking process with vibration; G) Water conduction height vs time for a root in a vibrating and non-vibrating container; Measurement of the meniscus angle for H) static and I) vibrating evaporator setup.
}
  \label{fig7}
\end{figure}

Vibrations in wastewater bodies are common, often arising from a combination of mechanical, hydraulic, biological, and external sources. It is necessary to evaluate the influence of the vibrations on the evaporator setup and the evaporation efficiency. The influence of vibrations of on particle fouling at the root surface is evaluated. For a non-vibrating root, a cluster of particles adsorb on the cotton surface, as shown in Fig. \ref{fig7}(A). On introducing vibrations, the cluster desorbs, exhibiting a comparatively cleaner cotton root. Image analysis is performed to quantify the percentage particle fouling on the root surface. For the static root, an 85\% surface coverage is reported, 25\% more than a vibrating root (60\%) (Fig. \ref{fig7}(B)). The vibrational forces act as a resistance to the adhesion between the particles and the cotton fibers, resulting in lower accumulation of particles on the surface.

To confirm the presence of silica particles, hand-held Raman spectroscopy is done on the surface of the root. A strong presence of silica particles is reported on both the static and the vibrating root, as shown in Fig. \ref{fig7}(C). The intensity of the silica particles for the static peak, however, is around 15\% higher than that for the vibrating root. This conforms well to our surface coverage results reporting more silica particle fouling for the static root. 

The influence of root fouling on the ER is evaluated with a SE setup under a wind velocity of 3.5 m/s, comparing a clean root with a fouled root for a 60 wt\% silica slurry. A 40\% decrease in ER is observed for the fouled root setup, decreasing from 3.9 kg/(m\(^{2}\)h for a cleaned root to 2.4 kg/(m\(^{2}\)h for a fouled root setup (Fig. \ref{fig7}(D)). This decrease is attributed to the close packing of the particles on the root surface, reducing the permeability and providing resistance to the entry of water to the root surface. The lower water absorption leads to a slower water conduction, resulting in decrease in the evaporation efficiency.

We further evaluate the influence of vibrations on the ER by introducing vibrations to the entire SE setup. The evaporation performance is assessed against 65 wt\% silica suspension and 60 wt\% tailing wastewaters under a wind velocity of 3.5 m/s, along with a control (natural evaporation). The ER with the SE setup is more than 4 times higher than the control (0.8 kg/(m\(^{2}\)h for silica, 0.5 kg/(m\(^{2}\)h for tailings), irrespective of the vibrations (Fig. \ref{fig7}(E)). However, the ER of the setups is similar, regardless of vibrations. This result is contrary to the silica fouling result discussed previously, where a reduction in particle adherence is observed with the introduction of vibrations.

To investigate, vertical water conduction tests is performed with the introduction of vibrations to the water container, as exhibited in Fig. \ref{fig7}(F). A lower vertical height is observed for the vibrating container, reducing to 5.5 cm from 6.5 cm for the non-vibrating container (Fig. \ref{fig7}(G)). This suggests the water conduction process is getting disrupted by the vibrations, which may lead to a slower water rise. For the water process, the capillary rise (h\textsubscript{cap}) within the tube is described by the formula:

\begin{equation}
h_{cap} = 2\sigma cos\theta/\rho gr
\end{equation}

where \(\sigma\) is the surface tension, \(\theta\) is the angle of meniscus and r is the radii of the capillary. For two setups utilizing cotton cloth for water conduction, vibrations may have an influence only on the angle of meniscus:

\begin{equation}
h_{cap} \propto cos\theta
\end{equation}

To verify, we utilize contact angle meter to measure the angle of meniscus for a static capillary against a vibrating capillary. Glass capillaries are filled with water and optical images are analyzed. For a static capillary, a meniscus angel of 42\(\degree\) was observed (Fig. \ref{fig7}(H)). Conversely, for a vibrating capillary, a higher angle of meniscus is reported, increasing by 6.5\(\degree\) to 48.5\(\degree\), with an error of 1.5\(\degree\) (Fig. \ref{fig7}(I)). The error for the vibrating capillary is due to the repeats, on account of the vibration-induced blurriness in the pictures. This slightly increased angle suggest a relatively lower capillary force in the capillary due to the vibrations, conforming well with our water conduction and evaporation results.

\subsection{Large-scale and outdoor evaporation experiments}

The wastewater within tailing ponds usually separate out into a top supernatant water layer and a bottom thicker tailing layer. While it is easy to remove the supernatant water, the real challenge lies in extracting water from within the tailing suspension. It is desirable to test and utilize long roots for evaporation at high solid concentration  \cite{zhang2022experimental,tsotsas2000drying}, without extracting water from the supernatant.

\begin{figure}[H]
  \centering
  \includegraphics [width = 16.5 cm]{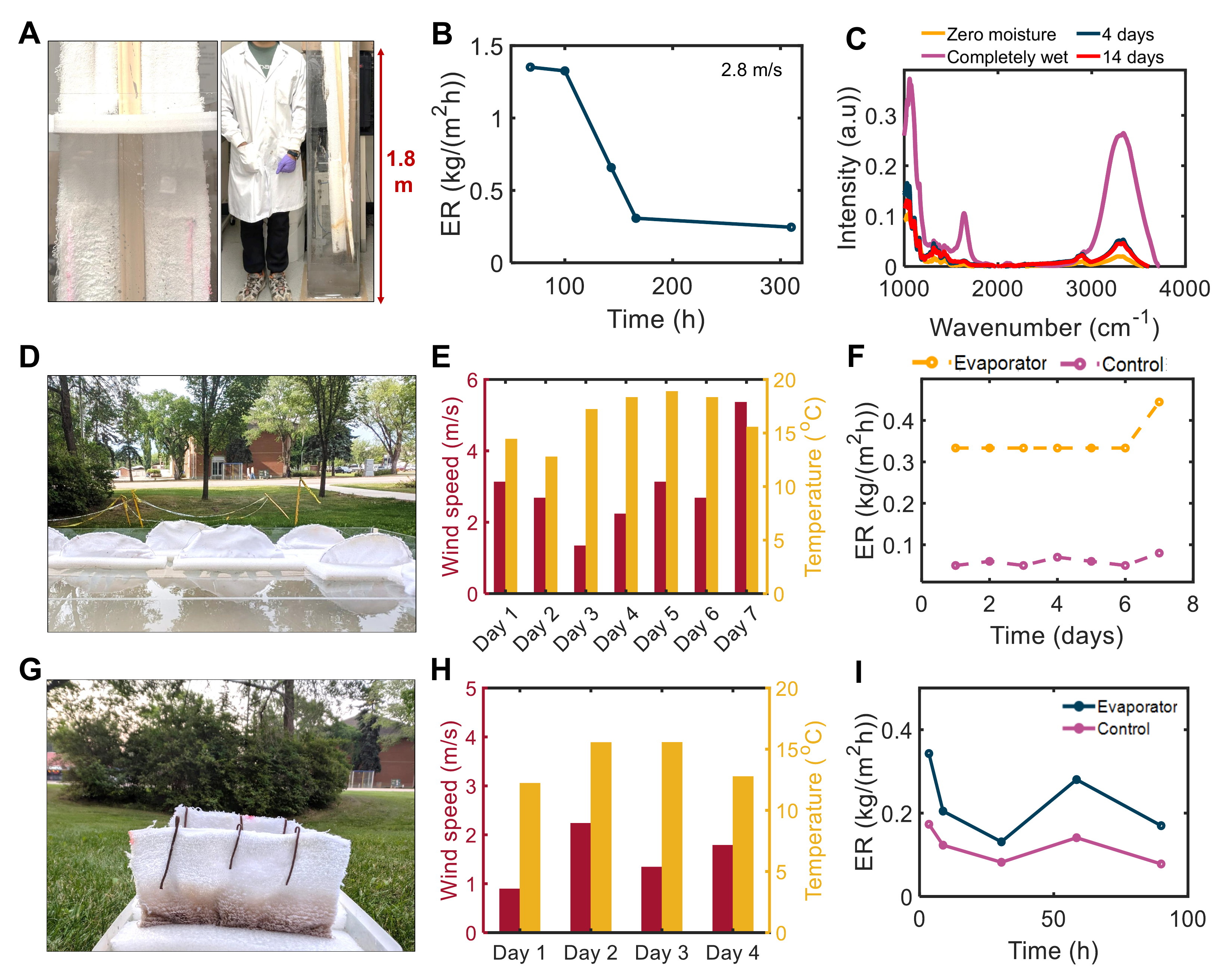}
  \caption{A) Image of the long root in the stratified SE setup; B) ER vs time for the stratified SE setup at 2.8 m/s; C) FTIR spectrum of water on the cotton sail surface after 4 and 14 days; D) Image of multiple SE setups in outdoor conditions; E) Average wind velocity and temperature on the 7 days of evaporation; F) ER vs time for the SE setups and control (natural evaporation); G) Image of the SE setup for tailing evaporation; H) Average wind velocity and temperature on the 4 days of evaporation; I) ER vs time for the SE setups and control (natural evaporation). }
  \label{fig8}
\end{figure}

As shown in Fig. \ref{fig8}(A), a SE setup with a pre-wetted 150 m long root is tested against 40 wt\% tailings under a wind velocity of 2.8 m/s. An ER of 1.5 kg/(m\(^{2}\)h is observed for the initial 100 hours, beyond which it drops below 0.5 kg/(m\(^{2}\)h up until 14 days of experiment (Fig. \ref{fig8}(B)). This sharp decrease in ER is attributed to minimal water conduction to the sail surface, as the height may be beyond the maximum capillary conduction height, even after including evapo-transpiration effects. To confirm, FTIR tests is performed to detect water on the sail surface. After 4 days of evaporation, only 22\% water is present on the sail surface, which further goes down to 17\% after 14 days of evaporation (Fig. \ref{fig8}(C)). This aptly explains the low ER observed by the stratified setup.

The practical application of the SE setups was demonstrated by conducting outdoor experiments with a control (natural evaporation) evaporating a 100 L container of water (Fig. \ref{fig8}(D)). An average temperature of $^{o}$C and wind speed of was noted throughout the experiment duration, as reported in Fig. \ref{fig8}(E). The evaporator displays nearly 8 times higher ER than control for nearly 8 days of the experiment, showcasing its superior evaporation performance to  natural evaporation (Fig. \ref{fig8}(F)).

The evaporation rate (ER) was further assessed using 60 wt\% fluid fine tailings over a four-day period and compared with natural evaporation conditions (Fig. \ref{fig8}(G)). The average wind velocity and ambient temperature during these days are illustrated in Fig. \ref{fig8}(H). The SE setup displayed close to twice the ER than natural evaporation under low wind conditions (Fig. \ref{fig8}(I)). These findings validate the potential of SE evaporators not only for large-scale slurry evaporation, but also its effectiveness in weak wind conditions, making it a practical, all-weather solution. Additionally, with the increased root depth enabling effective dewatering of tailings ponds, the setup displays immense potential in the practical treatment of industrial wastewaters, including the tailing ponds.

\section {Conclusions}

In this work, we demonstrated how to enhance the evaporation performance of a sailboat evaporator (SE) system for highly concentrated wastewaters. Beyond 75 wt\% solid concentration, a 33\% improvement in ER is observed with a dendritic root setup, attributed to wider spatial distribution of root material within the slurry. Using replantation of the evaporator, the SE achieved an evaporation rate (ER) of 4 kg/(m\(^{2}\)h) at 80 wt\% solids, representing a 25~\% improvement over non-replanted controls. This is reasoned to the re-established access to disconnected water pockets, as confirmed by reduced variance in spatial moisture distribution, maintaining uniform moisture gradients. Hydrodynamic flushing of the roots further enhanced ER by 75~\%, due to removal of silica accumulation near the evaporator root. This enhancement was supported by a theoretical inequality relating flushing forces and particle adhesion, presenting dependance on flushing fluid composition, flow velocity, and local solid packing density around the root volume. Mechanical vibrational inputs, while increasing resistance to particle adhesion on cotton fibers, failed to improve evaporation efficiency, likely due to reduced vertical water conduction and increased meniscus angle. Large-scale, stratified evaporation tests with 1.5~m roots demonstrated the system's larger-than-lab scalability under industrially relevant conditions. Outdoor experiments validated the SE system's robustness and all-weather applicability, highlighting its potential for practical deployment in industrial slurry treatment, including tailing ponds.

\section{Acknowledgments}

We are grateful for inspiring discussions with James Lockart and Mike Gattrell at BC Research throughout the entire program. We also thank Dr. Louis Kabwe for his guidance in the mechanical characterization and hydraulic conductivity experiments. We thank Yuki Maeda, Jason Chen and the team at Old Scona Academic School for assisting in fabrication of large evaporators and in performing the outdoor tests. The authors acknowledge the funding support from Mitacs Accelerate Program, the Canada Foundation for Innovation (CFI), the Natural Science and Engineering Research Council of Canada (NSERC). This work is partially supported by the Canada Research Chairs program (Award number: CRC-2018-00344). The authors acknowledge the assistance of DeepSeek (https://www.deepseek.com) for language refinement during the preparation of this manuscript.
After using this tool, the authors reviewed and edited the content thoroughly and take full responsibility for the content.

\bibliography{ref.bib}

\end{document}